\documentclass[aps,pra,showpacs,twocolumn,superscriptaddress,amsmath,amssymb]{revtex4}

\usepackage{graphicx,dcolumn,bm}
\usepackage{amsmath}
\usepackage{amsthm}
\usepackage{amsfonts}
\usepackage{dsfont}

\begin{document}

\title{Non-locality from N $>$ 2 Independent Single Photon Emitters}

\author{C. Thiel}
\affiliation{Institut f\"ur Optik, Information und Photonik, Universit\"at Erlangen-N\"urnberg, 91058 Erlangen, Germany}

\author{R. Wiegner}
\email{Ralph.Wiegner@physik.uni-erlangen.de}
\affiliation{Institut f\"ur Optik, Information und Photonik, Universit\"at Erlangen-N\"urnberg, 91058 Erlangen, Germany}

\author{J. von Zanthier}
\affiliation{Institut f\"ur Optik, Information und Photonik, Universit\"at Erlangen-N\"urnberg, 91058 Erlangen, Germany}

\author{G. S. Agarwal}
\affiliation{Department of Physics, Oklahoma State University, Stillwater, Oklahoma 74078-3072, USA}

\date{\today}

\begin{abstract}
We demonstrate that intensity correlations of second order in the fluorescence light of N~$>$~2 single-photon emitters may violate locality while the visibility of the signal remains below $1/\sqrt{2}\approx71\%$. For this, we derive a homogeneous Bell-Wigner-type inequality, which can be applied to a broad class of experimental setups. We trace the violation of this inequality back to path entanglement created by the process of detection.
\end{abstract}


\maketitle

\section{Introduction\label{1}}

The demonstration of non-locality for a system of more than two particles or even for an EPR state of two particles with higher spins has been an outstanding problem in the foundations of quantum theory. There are seminal papers on the theoretical aspects of this subject, notably from Mermin and others~\cite{Mermin:1980,Mermin:1982,Mermin:1990,Agarwal:1993,Zukowski:1993:a}. 
These papers pointed out the necessity for the use of other Bell-type inequalities \cite{Bell:1964:a} than the celebrated CHSH or CH74 inequalities \cite{Clauser:1969:a,Clauser:1974:a} in case of a system with N~$>$~2 particles or for higher spin systems, involving the measurement of N-particle properties. However, so far, experimental realizations of these inequalities do not seem to exist. 

In this paper we propose as a source for N~$>$~2 particles a chain of N independent single photon emitters, say trapped ions, trapped neutral atoms, quantum dots, or any other equivalent physical system with access to similar behavior, in order to test non-locality with more than two particles. Using this source, path entanglement among the emitted photons is created in the process of detection due to the absence of which-way information when registering a photon in the far field of the source~\cite{Maser:2009:a,Wiegner:2010:a}. Employing \textit{two} point photon-photon correlations denoted by $G^{(2)}({{\boldsymbol r}}_1, {{\boldsymbol r}}_2)$ in Glauber's notation~\cite{Glauber:1963:a} - a quantity which is easily accessible experimentally and has become the workhorse of experimentalists in quantum information science~\cite{Zeilinger:2008:a} - we are able to show that it is possible to violate a new form of Bell-type inequality. In particular, we show that the violation subsists for increasing N when the visibility of the photon-photon correlation signal continuously reduces to 33\%.
We note that several theoretical and experimental papers have dealt with the question of violations of Bell's inequalities employing the photon-photon correlation function for \textit{two} independent emitters~\cite{Mandel:1983:a,Ou:1988:a,Zukowski:1993:b}. In this case violation of locality has been demonstrated if the visibility of the two-photon 
signal exceeds $1/\sqrt{2}\approx71\%$, what is recovered by our results. Our work therefore clearly suggests that without indicating N the magnitude of the visibility of the G$^{(2)}$-signal can not be taken as a signature of non-locality.

In the following we start to introduce our system of N independent single photon emitters. Hereby we assume that each emitter is initially prepared in an excited state which has an appropriate Zeeman degeneracy so that each emitter upon spontaneous decay scatters either a right hand or a left hand circularly polarized photon. Alternatively, we could  consider $N$ two-level atoms with a $\lambda/4$ wave plate positioned in front turning the polarization of the scattered photon into the desired circular polarization. We further assume two spatially separated detectors in the far field of the source each capable of measuring single photon events. The two detectors are used to measure $G^{(2)}({{\boldsymbol r}}_1, {{\boldsymbol r}}_2)$ in order to characterize the two point correlations of the photons emitted by our source. As will be shown below, for N~$>$~2 emitters we need new Bell's inequalities to reveal the non-local behavior of the photon correlations. We derive these inequalities and obtain conditions on the spatial locations of the detectors in order to prove the non-local character of the photon correlations. These arise from the path entanglement created a posteriori by the selection of modes due to the process of detection~\cite{Wiegner:2010:a}. This is a novel aspect of our scheme as we do not need to start with a source producing entanglement ab initio among the N emitters.

The paper is organized as follows: in Sec.~\ref{2} we introduce our light source of $N$ uncorrelated single-photon emitters and explain how to describe a joint detection measurement of two photons in the far-field of this source. In Sec.~\ref{3}, we recapitulate the well-known set of CH74 inequalities~\cite{Clauser:1969:a,Clauser:1974:a} and explain  how these can be violated by the probability of finding two photons at two positions scattered by our source with N = 2 emitters and cannot be violated in case of N~$>$~2. In Sec.~\ref{4}, we  derive a new Bell-type inequality~\cite{Janssen:2004:a,Wildfeuer:2008:a} which allows to prove the non-local character of the correlations among the photons scattered by our source for any N~$\geq$~2. In Sec.~\ref{5} we finally conclude.

\section{Description of the physical system\label{2}}
\subsection{Setup of $N$ single photon emitters\label{21}}

\begin{figure}[h!]
\centering
\includegraphics[width=0.48\textwidth, bb=90 590 675 815]{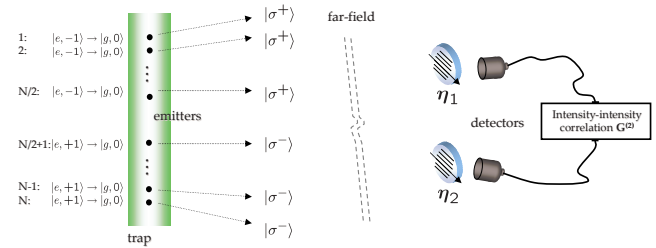}
\caption{\label{f2}Setup used for measuring the intensity correlation function of second order for a source of $N$ single-photon emitters of known origin and polarization. The photon detectors are placed at ${\bf r}_1$ and ${\bf r}_2$ in the far-field region of the emitters and are equipped with polarization filters in front, transmitting ${\boldsymbol\eta}_1$ and ${\boldsymbol\eta}_2$ polarized light, respectively.}
\end{figure}

We consider the setup shown in Fig.~\ref{f2}: $N$ single-photon emitters regularly arranged in a row at positions ${\bf R}_1, {\bf R}_2, \ldots, {\bf R}_N$ serve as a source for $N$ photons. The internal level scheme of the emitters is assumed to be characterized by a $V$-configuration, e.g., Zeeman sub-levels with two excited states $|e,-1\rangle$ and $|e,+1\rangle$, which both decay to a common ground state $|g,0\rangle$, accompanied by the emission of a ${\boldsymbol\sigma}^+$ or ${\boldsymbol\sigma}^-$ polarized photon, respectively. For the sake of simplicity we suppose that both transitions are equally probable. 
We further assume that for an even number of emitters, the first $N/2$ atoms are initially in the state $|e,-1\rangle$ and the remaining $N/2$ atoms in the state $|e,+1\rangle$. The initial state of the system can thus be written in the form
\begin{eqnarray}\label{e551}
|\psi_i\rangle = \prod_{n=1}^{N/2}|e,-1\rangle_n\otimes\prod_{n=\frac{N+2}{2}}^{N}|e,+1\rangle_n,
\end{eqnarray}
where the subscripts $n$ refers to the atom located at ${\bf R}_n$. For an odd number of emitters, we suppose that the first $(N-1)/2$ emitters are initially in $|e,-1\rangle$ and the remaining $(N+1)/2$ emitters in $|e,+1\rangle$ so that the initial state is given by
\begin{eqnarray}\label{e552}
|\psi_i\rangle = \prod_{n=1}^{\frac{N-1}{2}}|e,-1\rangle_n\otimes\prod_{\frac{N+1}{2}}^{N}|e,+1\rangle_n.
\end{eqnarray}

Due to the process of spontaneous decay the $N$ three-level emitters will scatter exactly $\frac{N}{2}$ ($\frac{N-1}{2}$) ${\boldsymbol\sigma}^+$ and $\frac{N}{2}$ ($\frac{N+1}{2}$) ${\boldsymbol\sigma}^-$ polarized photons in the case of an even (odd) number of emitters. Alternatively, we could also consider $N$ two-level atoms with $\lambda/4$ wave plates positioned in front of each particle which turn the polarization of the photons emitted by the atom at ${\bf R}_n$ for $n = 1 \ldots \frac{N}{2} \,  (\frac{N-1}{2})$ into ${\boldsymbol\sigma}^+$ and for $n = \frac{N}{2} \, (\frac{N+1}{2}) \ldots N$ into ${\boldsymbol\sigma}^-$ polarization in the case of an even (odd) number of emitters; the only prerequisite for our scheme is that a precisely determined number of ${\boldsymbol\sigma}^+$ and ${\boldsymbol\sigma}^-$ polarized photons of known origin is emitted by the setup. 

In order to measure the intensity correlation function of second order we locate two detectors at ${\bf r}_1$ and ${\bf r}_2$ in the the far-field region of the emitters, each equipped with a polarization filter in front, oriented along ${\boldsymbol\eta}_1$ and ${\boldsymbol\eta}_2$, respectively. The operator $\hat{D}_N(\delta({\bf r}_j),{\boldsymbol\eta}_j)$ which describes a successful detection event of a photon at the detector at ${\bf r}_j$ ($j=1,2$), after having passed a polarization filter oriented along ${\boldsymbol\eta}_j=\sin\vartheta_j{\boldsymbol\sigma}^++\cos\vartheta_j{\boldsymbol\sigma}^-$,  can be written in case of an even number of emitters in the initial state~(\ref{e551}) in the form~\cite{Thiel:2007:a}
\begin{eqnarray}\label{e553}
\hat{D}_N(\delta({\bf r}_j),{\boldsymbol\eta}_j)=\frac{E_0}{\sqrt{2}} (\sin{\vartheta_j}\sum\limits_{n=1}^{N/2}e^{in\delta_j}|g,0\rangle_n\langle e,-1|+\nonumber\\
+\cos{\vartheta_j}\!\!\!\sum\limits_{n=\frac{N+2}{2}}^{N}\!\!\!e^{in\delta_j}|g,0\rangle_n\langle e,+1|)\vspace{2cm}
\end{eqnarray}
where the sum over $n$ takes into account that principally each atom could have emitted the recorded photon. 
Here, $|g,0\rangle\langle e,\pm1|$ is an atomic operator projecting the atomic state $|e,\pm1\rangle$ onto $|g,0\rangle$, $E_0$ is the amplitude of the electric field, and $\delta({\bf r}_j)$ is the optical phase difference between photons being emitted by adjacent atoms and registered at ${\bf r}_j$. In the far-field and using a coordinate system where ${\bf R}_n=n\,{\bf R}_1$ (with ${\bf R}_0\equiv{\bf 0}$) the optical phase difference $\delta_j$ is given by~\cite{Thiel:2007:a}
\begin{eqnarray}\label{e23}
\delta({\bf r}_j):= \delta_j = kd\,\frac{{\bf r}_j\cdot{\bf R}_1}{|{\bf r}_j|\,|{\bf R}_1|}= kd\,\sin\theta_j,
\end{eqnarray}
where $k$ denotes the wavenumber of the scattered light, $d$ the interatomic spacing and $\theta_j$ the scattering angle as shown in Fig.~\ref{f2}. In analogy, the detection operator for an odd number of $N$ emitters acting on the initial state~(\ref{e552}) can be written as
\begin{widetext}
\begin{eqnarray}\label{e554}
\hat{D}_N({\delta}_j,{\boldsymbol\eta}_j)=\frac{E_0}{\sqrt{2}}\left(\sin{\vartheta_j}\sum\limits_{n=1}^{\frac{N-1}{2}}e^{in\delta_j}|g,0\rangle_n\langle e,-1|+\cos{\vartheta_j}\!\!\!\sum\limits_{n=\frac{N+1}{2}}^{N}\!\!\!e^{in\delta_j}|g,0\rangle_n\langle e,+1|\right).
\end{eqnarray}

\subsection{Intensity correlation signal of second order\label{22}}

With the detection operators $\hat{D}_N({\delta}_j,{\boldsymbol\eta}_j)$ at hand, we can calculate from Eqs.~(\ref{e551})-(\ref{e554}) the intensity correlation function of second order $G^{(2)}_N(\delta_1,\delta_2;{\boldsymbol\eta}_1,{\boldsymbol\eta}_2)$ for our system of $N$ single photon emitters. Hereby, we assume in the following that the first two out of $N$ scattered photons are recorded by the two detectors~\footnote{If other photons than the first two are measured the expressions for $G^{(2)}_N$ have to be multiplied by an overall factor. This does not change the visibility of the expressions but introduces different normalization factors when considering successive measurements. 
More information regarding this topic is given at the end of the current Section.}. 
According to~\cite{Thiel:2007:a} we then have
\begin{eqnarray}\label{e532}
G^{(2)}_N(\delta_1,\delta_2;{{\boldsymbol\eta}}_1,{{\boldsymbol\eta}}_2):= \left|\hat{D}_N(\delta_2,{\boldsymbol\eta}_2)\,\hat{D}_N(\delta_1,{\boldsymbol\eta}_1)|\psi_i\rangle\right|^2.
\end{eqnarray}
As there is a unique correspondence between ${\boldsymbol\eta}_j$ and $\vartheta_j$, we can write in the following  $G^{(2)}_N(\delta_1,\delta_2;{\boldsymbol\eta}_1,{\boldsymbol\eta}_2)$ also as $G^{(2)}_N(\delta_1,\delta_2;\vartheta_1,\vartheta_2)$.
By fixing the orientation of the polarization filters in front of the two detectors identical to $\vartheta_1=\vartheta_2=\frac{\pi}{4}$, corresponding to ${\boldsymbol\eta}_j=1/\sqrt{2}({\boldsymbol\sigma}^++{\boldsymbol\sigma}^-)$ (for $j=1,2$), we obtain
\begin{eqnarray}\label{e555}
G^{(2)}_N(\delta_1,\delta_2;\frac{\pi}{4},\frac{\pi}{4})=\frac{E_0^4}{8}\,\left(1+\frac{2}{N\,(N-1)}\,\sum\limits_{n=1}^N(N-n)\,\cos(n\,(\delta_2-\delta_1))\right),
\end{eqnarray}
which holds for even or odd $N$. This function is illustrated for different $N$ in Fig.~\ref{f55}.
\begin{figure}[t!]
\centering
\includegraphics[width=0.7\textwidth, bb=60 440 760 815, clip=true]{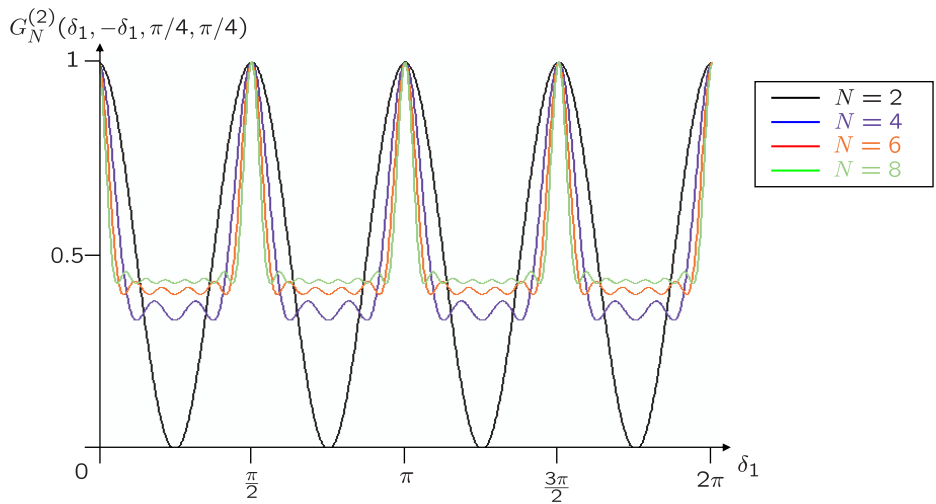}
\caption{\label{f55}Plot of the intensity correlation function of second order $G^{(2)}_N(\delta_1,-\delta_1,\frac{\pi}{4},\frac{\pi}{4})$ (in arbitrary units) as a function of the relative phase shift $\delta_1$  (c.f.~Eq.~(\ref{e555})). The plot illustrates the signal for different numbers of emitters $N$ ($N=2,4,6,8$).}
\end{figure}
In the case that the two polarizers are set orthogonal at $\vartheta_1=\frac{\pi}{4}$ and $\vartheta_2=\frac{3\pi}{4}$ (corresponding to ${\boldsymbol\eta}_1=1/\sqrt{2}({\boldsymbol\sigma}^++{\boldsymbol\sigma}^-)$ and ${\boldsymbol\eta}_2=1/\sqrt{2}({-\boldsymbol\sigma}^++{\boldsymbol\sigma}^-)$, respectively) we find for an even number of emitters $N$
\begin{eqnarray}\label{e556}
G^{(2)}_N(\delta_1,\delta_2;\frac{\pi}{4},\frac{3\pi}{4})&=&\frac{E_0^4}{8}\,\left(1+\frac{2}{N\,(N-1)}\,\sum\limits_{n=1}^{N/2}(N-2n)\,\cos(n\,(\delta_2-\delta_1))\right.\\
&-&\left.\frac{2}{N\,(N-1)}\,\sum\limits_{\alpha=1}^{N/2}\sum\limits_{n=1}^{N}\left(\Theta(N-n-\alpha+1)\,\cos(n\,(\delta_2-\delta_1))\,\Theta(n-\alpha+1)\right)\right),\nonumber
\end{eqnarray}
where the Heaviside step function $\Theta(x)$ is defined as 
\begin{eqnarray}
\Theta(x):=\left\{\begin{array}{cc}
0&x\leq0\\
1&x>0
\end{array}\right..
\end{eqnarray}
In analogy, we find for an odd number of emitters $N$
\begin{eqnarray}\label{e557}
G^{(2)}_N(\delta_1,\delta_2;\frac{\pi}{4},\frac{3\pi}{4})&=&\frac{E_0^4}{8}\,\left(1+\frac{2}{N\,(N-1)}\,\sum\limits_{n=1}^{\frac{N-1}{2}}(N-2n)\,\cos(n\,(\delta_2-\delta_1))\right.\\
&-&\left.\frac{2}{N\,(N-1)}\,\sum\limits_{\alpha=1}^{\frac{N-1}{2}}\sum\limits_{n=1}^{N}\left(\Theta(N-n-\alpha+1)\,\cos(n\,(\delta_2-\delta_1))\,\Theta(n-\alpha+1)\right)\right).\nonumber
\end{eqnarray}

\end{widetext}
From Eq.~(\ref{e555}) we can calculate the visibility ${\cal V}_N$ of the intensity correlation signal of second order $G^{(2)}_N(\delta_1,\delta_2;\frac{\pi}{4},\frac{\pi}{4})$ in case of identically oriented polarizers. 
For even or odd $N$ we find 
\begin{eqnarray}\label{VN}
{\cal V}_N:=\frac{max[G^{(2)}_N]-min[G^{(2)}_N]}{max[G^{(2)}_N]+min[G^{(2)}_N]}=\frac{N}{3\,N-4},
\end{eqnarray}
where $max[G^{(2)}_N]$ ($min[G^{(2)}_N]$) corresponds to the maximum (minimum) value of the function $G^{(2)}_N\equiv G^{(2)}_N(\delta_1,\delta_2;\frac{\pi}{4},\frac{\pi}{4})$. Eq.~(\ref{VN}) shows that ${\cal V}_N$ can be uniquely assigned to the number of emitters $N$. 
Note that ${\cal V}_N$ represents an ideal theoretical value only; in general, experimental uncertainties and insufficiencies will influence and decrease the attainable visibility.\\

In the derivation of Eqs.~(\ref{e555})-(\ref{VN}) it has been assumed that the two photons measured are the first two photons being scattered by our system of $N$ single photon emitters. This scenario enables to work with the initial states given by Eqs.~(\ref{e551}) or (\ref{e552}). One way to achieve this experimentally is to measure all $N$ scattered photons and pick out the first two detection events via post-selection. Thereby the experimental challenge of measuring the intensity correlation function of second order for $N$ possible emitters appears to equal the requirements of measuring the intensity correlation function of $N$th order (see, e.g.,~\cite{Thiel:2007:a}). We note, however, that our measurement scheme requires to resolve the spatial distribution of the two-photon correlation signal $G^{(2)}_N(\delta_1,\delta_2;\vartheta_1,\vartheta_2)$ only and that one can make use of a large bucket detector or a lens system to detect the remaining $N-2$ photons what simplifies the requirements.

Besides experimental challenges, the restriction of detecting the first two photons bears a major advantage: since our system consists of a fixed number of scatterers, the number of photons contributing to a successful measurement cycle is precisely known. Therefore, the intensity correlation function of $N$th order is directly proportional to the probability of finding $N$ photons. In particular, the intensity correlation signal of second order $G^{(2)}_N(\delta_1,\delta_2;\vartheta_1,\vartheta_2)$ is related to the detection probability of finding jointly the first two photons $p_{12}^N(\delta_1,\delta_2;\vartheta_1,\vartheta_2)$ via
\begin{eqnarray}\label{probability}
p_{12}^N(\delta_1,\delta_2;\vartheta_1,\vartheta_2)=\frac{{\cal C}_0^2}{{E_0^4}}\,G^{(2)}_N(\delta_1,\delta_2;\vartheta_1,\vartheta_2),
\end{eqnarray}
where the superscript $N$ denotes the number of emitters used in the setup and ${\cal C}_0:=\mu\,\frac{\Delta\Omega}{4\pi}$ abbreviates the overall success probability to find a single photon at a detector with quantum efficiency $\mu$ and subtending a solid angle $\Delta\Omega$.

\section{CH74 inequalities for multiple emitters\label{3}}

In his seminal paper Bell proved that deterministic local theories with hidden variables are incompatible with quantum mechanics~\cite{Bell:1964:a}. In this Section, we want to apply this criterium to investigate whether the photons emitted by our system of $N$ regularly arranged single photon emitters display spatial correlations which are compatible or incompatible with local deterministic theories. For this, we recapitulate a well-known set of homogeneous position dependent Bell-type inequalities, the so-called CH74 inequalities~\cite{Clauser:1974:a,Clauser:1969:a,Ou:1988:a}, which we then apply for our system of $N$ single-photon emitters.

\subsection{Theory of CH74 inequalities\label{31}}

Let us denote the continuous set of hidden variables by $\lambda$. The probability of registering one photon out of a set of $N$ single photon emitters at a position ${\bf r}_j$ is then determined by $p^N({\bf r}_j,\lambda)$, where we included the hidden variables $\lambda$ in the argument of the single photon detection probability $p^N({\bf r}_j)$. Following the requirement of {\em locality}, the joint probability $p^N_{12}({\bf r}_1,{\bf r}_2,\lambda)$ of detecting two photons at ${\bf r}_1$ and ${\bf r}_2$ can be written as the product of the two independent single detection probabilities
\begin{eqnarray}\label{pro0}
p_{12}^N({\bf r}_1,{\bf r}_2,\lambda)=p^N({\bf r}_1,\lambda)\cdot p^N({\bf r}_2,\lambda).
\end{eqnarray}
Though $\lambda$ are hidden variables of a deterministic local theory and thus unknown, the detection probabilities obtained when performing a real experiment are determined by the ensemble averages over all $\lambda$
\begin{eqnarray}\label{pro1}
&p^N({\bf r}_j)&=\int d\lambda\,g(\lambda)\,p^N({\bf r}_j,\lambda)\quad\mbox{with}\quad j=1,2\nonumber,\\
&p_{12}^N({\bf r}_1,{\bf r}_2)&=\int d\lambda\,g(\lambda)\,p^N({\bf r}_1,\lambda)\,p^N({\bf r}_2,\lambda),
\end{eqnarray}
where $g(\lambda)$ denotes an appropriate weight function of the hidden variables.

Having introduced the single photon and joint detection probabilty $p^N({\bf r}_j,\lambda)$ and $p_{12}^N({\bf r}_i,{\bf r}_j,\lambda)$, respectively, the homogeneous CHSH-type inequalities can be derived from the following mathematical inequalities~\cite{Clauser:1974:a},
\begin{eqnarray}\label{ch1}
-XY\leq x\,y-x\,y'+x'\,y+x'\,y'-Yx'-Xy\leq0,
\end{eqnarray}
These inequalities hold for any values $x,x',y,y',X,Y$ fulfilling $0\leq x,x'\leq X$ and $0\leq y,y'\leq Y$. Setting $X = Y = 1$, so that $0\leq x,x',y,y'\leq 1$, we can then identify
\begin{eqnarray}\label{set4}
&\hspace{-4mm}p^N(\delta_1,\vartheta_1,\lambda)\!=\!x,\;p^N(\delta_1',\vartheta_1,\lambda)\!=\!x',\;p^N(\delta_1,\infty,\lambda)\!=\!X,\nonumber\\
&\hspace{-6mm}p^N(\delta_2,\vartheta_2,\lambda)\!=\!y,\;p^N(\delta_2',\vartheta_2,\lambda)\!=\!y',\;p^N(\delta_2,\infty,\lambda)\!=\!Y.
\end{eqnarray}

where the arguments of the probabilities refer to our setup: the $j$th detector is sensitive to $\vartheta_j$ polarized light only and is located at ${\bf r}_j$ ($j=1,2$) where Eq.~(\ref{e23}) relates the detector position ${\bf r}_j$ to the optical phase $\delta_j$. The notation $\infty$ indicates that the polarization filter is removed for the particular measurement. The constraint $X\geq x,x'$ ($Y\geq y,y'$) is then guaranteed by the so-called {\em no-enhancement} condition~\cite{Clauser:1969:a,Clauser:1974:a,Ou:1988:a}: the detection probability when using a polarization filter cannot exceed a measurement without a polarization filter. Finally, in agreement with the requirements of a local hidden variable (LHV) theory and Eq.~(\ref{pro0}), we can write the two-photon joint detection probability as

\begin{eqnarray}\label{pro8}
p^N_{12}(\delta_1,\delta_2;\vartheta_1,\vartheta_2,\lambda)=p^N(\delta_1,\vartheta_1,\lambda)\cdot p^N(\delta_2,\vartheta_2,\lambda).
\end{eqnarray}

Combining Eqs.~(\ref{set4}) and (\ref{pro8}) with Eq.~(\ref{ch1}) we obtain, after multiplying the whole expression with $g(\lambda)$ and integrating over $\lambda$, the following inequality:
\begin{widetext}
\begin{eqnarray}\label{ch14b}
S_N\!&\!\!:=\!\!&\!\!\left[p^N_{12}(\delta_1,\delta_2;\vartheta_1,\vartheta_2)-p^N_{12}(\delta_1,\delta'_2;\vartheta_1,\vartheta_2)+p^N_{12}(\delta'_1,\delta_2;\vartheta_1,\vartheta_2)+p^N_{12}(\delta'_1,\delta'_2;\vartheta_1,\vartheta_2)\right.\!\nonumber\\
\!&\!\!-\!\!&\left.p^N_{12}(\delta'_1,\delta_2;\vartheta_1,\infty)-p^N_{12}(\delta_1,\delta_2;\infty,\vartheta_2)\right]/p^2_{12}(\delta_1,\delta_2;\infty,\infty)\leq0.
\end{eqnarray}
\end{widetext}
Hereby, we restricted ourselves to the upper bound of the inequalities~(\ref{ch1}) which allows to normalize the expression by an arbitrary function. In the following we choose as normalization function the expression $p^2_{12}(\delta_1,\delta_2;\infty,\infty)$ which is a constant, independent of $N$. This allows in particular for a better comparability of the results obtained in the forthcoming sections. 

Eq.~(\ref{ch14b}) is the  position dependent CHSH inequality which can be used to investigate the quantum nature of the spatial correlations of the photons emitted by our source depicted in Fig.~\ref{f2} (see also~\cite{Ou:1988:a,Wiegner:2010:a}). Note that, although we are interested in the spatial behavior of the two-photon correlation signal, the polarization degrees of freedom play a crucial role in the measurements of $p^N$ and $p^N_{12}$: we have to include them necessarily in order to satisfy the no-enhancement condition~\cite{Clauser:1969:a,Clauser:1974:a,Ou:1988:a}. We emphasize, however, that in our investigations we focus on the \textit{spatial} correlations among the emitted photons. 

In order to violate the inequality~(\ref{ch14b}) maximally, it is advantageous to adjust the polarization filters such that the detection efficiency of the experimental setup is optimized. In the following  we thus choose $\vartheta_1=\vartheta_2=\frac{\pi}{4}$ which yields the best results. With these settings, using Eq.~(\ref{probability}) and employing the relation $p_{12}(\delta_1,\delta_2;\vartheta_1,\infty)=p_{12}(\delta_1,\delta_2;\vartheta_1,\vartheta_2)+p_{12}(\delta_1,\delta_2;\vartheta_1,\vartheta_2+\frac{\pi}{2})$ we calculate the joint detection probabilities needed in Eq.~(\ref{ch14b}) to
\begin{eqnarray}
\label{pro7a}p^N_{12}(\delta_1,\delta_2;\frac{\pi}{4},\frac{\pi}{4})&=&\frac{{\cal C}_0^2}{E_0^4}\,G^2_N(\delta_1,\delta_2;\frac{\pi}{4},\frac{\pi}{4}),\\
\label{pro7b}p^N_{12}(\delta_1,\delta_2;\frac{\pi}{4},\infty)&=&\frac{{\cal C}_0^2}{E_0^4}\,G^2_N(\delta_1,\delta_2;\frac{\pi}{4},\frac{\pi}{4})\\
&+&\frac{{\cal C}_0^2}{E_0^4}\,G^2_N(\delta_1,\delta_2;\frac{\pi}{4},\frac{3\pi}{4}),\nonumber\\
\label{pro7c}p^2_{12}(\delta_1,\delta_2;\infty,\infty)&=&{\cal C}_0^2\,\frac{1}{2},
\end{eqnarray}
where we made use of the expressions derived in Eqs.~(\ref{e555}),~(\ref{e556}) and~(\ref{e557}). Whether or not a violation of the position dependent inequality Eq.~(\ref{ch14b}) for $N\geq2$ does occur can be verified by inserting Eqs.~(\ref{pro7a}) - (\ref{pro7c}) into Eq.~(\ref{ch14b}) and looking thereafter for the maxima of $S_N$ as a function of $\delta_1$, $\delta_2$, $\delta'_1$ and $\delta'_2$.

\subsection{CH74 inequalities for a system of two single-photon emitters (case N=2)\label{32}}

For the case of $N=2$ emitters, we find from Eqs.~(\ref{e555}) and (\ref{e556})
\begin{eqnarray}
\label{e538a}G^{(2)}_2(\delta_1,\delta_2;\frac{\pi}{4},\frac{\pi}{4})& = & \frac{E_0^4}{8}\,(1+\cos{[\delta_2-\delta_1]})\\
\label{e538b}G^{(2)}_2(\delta_1,\delta_2;\frac{\pi}{4},\frac{3\pi}{4})& = & \frac{E_0^4}{8}\,(1-\cos{[\delta_2-\delta_1]})
\end{eqnarray}
so that Eqs.~(\ref{pro7a}) - (\ref{pro7c}) become 
\begin{eqnarray}
\label{pro6a}p^2_{12}(\delta_1,\delta_2;\frac{\pi}{4},\frac{\pi}{4})&=&{\cal C}_0^2\,\frac{1}{8}(1+\cos{[\delta_2-\delta_1]}),\hspace{0.5cm}\\
\label{pro6b}p^2_{12}(\delta_1,\delta_2;\vartheta_1,\infty)&=&{\cal C}_0^2\,\frac{1}{4},\\
\label{pro6c}p^2_{12}(\delta_1,\delta_2;\infty,\infty)&=&{\cal C}_0^2\,\frac{1}{2}.
\end{eqnarray}
Plugging these results into Eq.~(\ref{ch14b}) we obtain
\begin{eqnarray}\label{ch11}
S_2 &=&\frac{1}{4}\Big(\cos{[\delta_2-\delta_1]}-\cos{[\delta_2'-\delta_1]}\\
&+&\cos{[\delta_2-\delta_1']}+\cos{[\delta_2'-\delta_1']}\Big)-\frac{1}{2}\leq0.\nonumber
\end{eqnarray}
Looking for the extrema of $S_2$ we find the following set of parameters (see also~\cite{Ou:1988:a,Wiegner:2010:a})
\begin{eqnarray}
\label{set1a}\delta_2-\delta_1=\frac{1}{8}\,2\pi, & \delta_2'-\delta_1=\frac{3}{8}\,2\pi,\vspace{0.1cm}\\
\delta_2-\delta_1'=\frac{1}{8}\,2\pi, & \delta_2'-\delta_1'=\frac{1}{8}\,2\pi, \nonumber
\end{eqnarray}
which lead, in combination with~(\ref{ch11}), to the following inequality with respect to the spatial correlations of the photons scattered by two single-photon emitters
\begin{eqnarray}\label{ch12}
S_2 = \sqrt{2}-1\leq0.
\end{eqnarray}

The inequality Eq.~(\ref{ch12}) is derived assuming an ideal visibility of 100\% for the two-photon correlation functions (Eqs.~(\ref{e538a}) and~(\ref{e538b})). However, the visibility that can be achieved in a real experiment is usually below that value due to experimental uncertainties, limited detector efficiencies etc.
Taking a reduced visibility ${\cal V} < 1$  for $G^{(2)}_2(\delta_1,\delta_2;\frac{\pi}{4},\frac{\pi}{4})$ and $G^{(2)}_2(\delta_1,\delta_2;\frac{\pi}{4},\frac{3\pi}{4})$ into account, Eq. (\ref{ch12}) reads:  
\begin{eqnarray}\label{ch12V}
S_2 =\sqrt{2} \cdot {\cal V} - 1\leq0.
\end{eqnarray}
This inequality may be violated only if the visibility exceeds $\frac{1}{\sqrt{2}}\approx71\%$~\cite{Ou:1988:a,Wiegner:2010:a,Zukowski:1993:b}. 

\subsection{CH74 inequalities for a system of multiple single-photon emitters (case $N>2$)\label{33}}

For the case of $N=2$ emitters, the extrema of $S_2$ are obtained using the set of analytical expressions for $\delta_1,\delta_1',\delta_2,\delta_2'$ provided in Eq.~(\ref{set1a}). These can also be written in the form 
\begin{eqnarray}\label{e559}
\delta_1=\alpha_1\,2\pi,\quad\delta_2=(\frac{1}{8}+\alpha_2)\,2\pi,\\
\delta_1'=(\frac{2}{8}+\alpha_3)\,\pi,\quad\delta_2'=(\frac{3}{8}+\alpha_4)\,\pi,\nonumber
\end{eqnarray}
with $\alpha_i \in {\mathbb N}$ ($i=1,...,4$).

In contrast, for $N>2$, the joint detection probabilities present in $S_N$ get more involved (c.f.~Eqs.~(\ref{e555}),~(\ref{e556}) and~(\ref{e557})). The values for $\delta_1,\delta_1',\delta_2,\delta_2'$ giving rise to maxima of $S_N$ were thus determined numerically. This approach unveiled that the maxima of $S_N$ for any $N>2$ can be obtained by choosing
\begin{eqnarray}\label{e558}
\delta_1=\alpha_1\,2\pi,\quad\delta_2=\alpha_2\,2\pi,\quad\delta_1'=\alpha_3\,\pi,\quad\delta_2'=\alpha_4\,\pi,
\end{eqnarray}
again with $\alpha_i \in {\mathbb N}$ ($i=1,...,4$, $\alpha_3,\alpha_4\not=0$).

\begin{figure}[t!]
\centering
\includegraphics[width=0.45\textwidth, bb=65 420 500 790, clip=true]{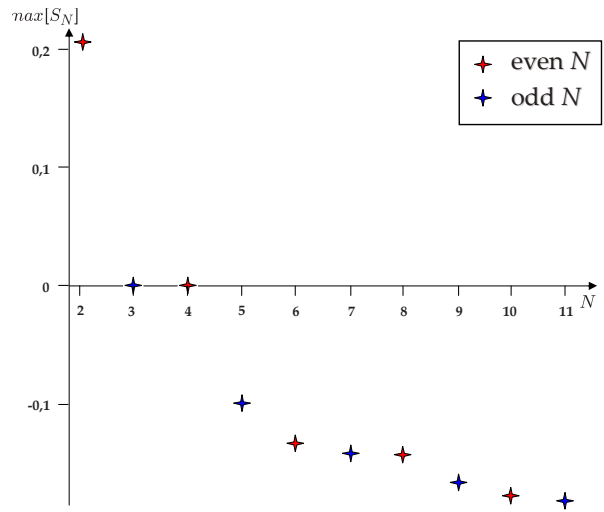}
\caption{\label{f56}Maximum values of $S_N$ as a function of the number of emitters $N$: the numerical calculations show that the CHSH-type inequality~(\ref{ch14b}) is violated only for $N=2$, whereas for $N>2$ the inequality holds. For $N=3$ and $N=4$ the maximum value corresponds to $S_{3,4}=0$.}
\end{figure}

The results of our numerical calculations for the maxima of $S_N$ (for $N=2,...,10$) are shown in Fig.~\ref{f56}. For $N=2$ we obtain as before $S_2=\sqrt{2}-1$. For $N=3,4$ we find in both cases $S_3=S_4=0$. For $N>4$ the values of $S_N$ are displayed in the plot: we see that the behavior is slightly different for even $N$ (red stars) and for odd $N$ (blue stars). However, we find that a violation of $S_N$ appears only for the case $N=2$.

\begin{figure}[t!]
\centering
\includegraphics[width=0.45\textwidth, bb=70 400 485 795, clip=true]{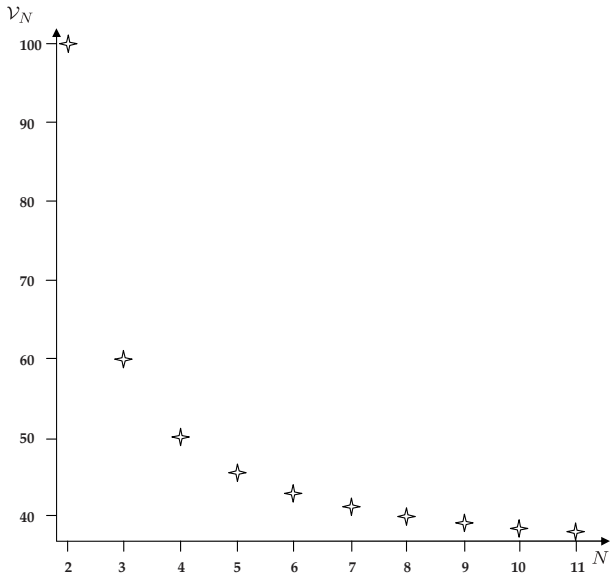}
\caption{\label{f57}Plot of the theoretical visibility ${\cal V}_N$ (c.f.~Eq.~(\ref{VN})) of the intensity correlation signal of second order as a function of the number of emitters $N$ for $N = 1, \ldots, 11$; the visibility reaches 50\% for $N=4$ and 33\% for ${\cal V}_N\rightarrow\infty$, respectively.}
\end{figure}

In conclusion, we see from Fig.~\ref{f56} that $S_N$ cannot be violated by the setup shown in Fig.~\ref{f2} for $N>2$ emitters. Taking into account the visibility of the intensity correlation function of second order as derived in Sec.~\ref{22} (c.f Eq.~(\ref{VN})) this result is in agreement with~Eq.(\ref{ch12V})~\cite{Ou:1988:a,Wiegner:2010:a}: while the joint detection probability for our setup in case of $N=2$ shows a modulation with a theoretical visibility of ${\cal V}_2=100\%$, Eq.~(\ref{VN}) reveals that ${\cal V}_N$ drops rapidly with $N>2$. This is illustrated in Fig.~\ref{f57}: already for the case of $N=3$ the visibility is reduced to ${\cal V}_3=60\%$, i.e., below the critical value of $1/\sqrt{2}\approx71\%$. The latter was found to be the required value in order to violate the Bell-type inequalities (\cite{Clauser:1969:a,Clauser:1974:a}, c.f.~Eq.(\ref{ch12V})).


Triggered by these results, we will consider a different inequality which is more appropriate for our system in the next Section. As it turns out this inequality is able to prove that the spatial intensity-intensity correlations of the photons spontaneously emitted by our source of $N$ single photon emitters are non-local in nature, even in the case of $N>2$ emitters, i.e., for a visibility ${\cal V}_N < 71\%$.

\section{A more suitable inequality for multiple emitters\label{4}}

In the following we introduce a new Bell-type inequality which allows to reveal the non-classical nature of the spatial intensity-intensity correlations even in the case that a two-photon correlation signal with a
visibility less than 71\% is measured. In fact, as will be shown, this new Bell-type inequality allows to reveal the non-classical character of the two-photon signal even for a visbility approaching 33\%. To demonstrate this, we start with a different mathematical inequality based on a so-called Bell Wigner-inequality (see, e.g.,~\cite{Pitowsky:1989:a,Janssen:2004:a}).

\subsection{Derivation of a homogeneous Bell-Wigner (HBW) inequality\label{41}}

The Bell-Wigner inequality can be written in the following form~\cite{Janssen:2004:a}
\begin{eqnarray}\label{pit}
0\leq x_1-x_1\,x_2-x_1\,x_3+x_2\,x_3,
\end{eqnarray}
which holds under the condition that $0\leq x_1,x_2,x_3\leq1$; for a proof of this inequality we refer to~\cite{Pitowsky:1989:a} (see also Appx.~\ref{app2}). By identifying $x_j$ ($j=1,2,3$) again with single photon detection probabilities we could speak of Eq.~(\ref{pit}) as an (inhomogeneous) Bell-type inequality since it considers both single photon \emph{and} joint detection probabilities. However, as motivated in the derivation of the CHSH-type inequality above, the experimental requirements can be eased if the inequality under investigation involves only detection probabilities of the same order. Hence, our goal is to derive a  \emph{homogeneous} Bell-type inequality on the basis of the above Bell-Wigner inequality which considers joint detection probabilities only, being subject to the same overall success probability. Our proposal for a new inequality reads
\begin{eqnarray}\label{BWT}
0\leq x_1\,x_4-x_1\,x_2-x_1\,x_3+x_2\,x_3,
\end{eqnarray}
and holds for the constraints $0\leq x_1,x_2,x_3\leq x_4\leq1$. The proof of~(\ref{BWT}) is provided in Appx.~\ref{app2}.

In analogy to the foregoing Section, we consider the setup with an even (odd) number of emitters $N$ as displayed in Fig.~\ref{f2}.
Again, a photon detection event registered at the $j$th detector is characterized by two parameters: the position ${\bf r}_j$ giving rise to an optical phase $\delta_j$ and the orientation of the $j$th polarizer ${\boldsymbol\eta}_j$ which we choose to be oriented along $1/\sqrt{2}({\boldsymbol\sigma}^-+{\boldsymbol\sigma}^+)$ ($j=1,2$), corresponding to $\vartheta_2=\vartheta_1=\frac{\pi}{4}$. The latter optimizes the overall success of the photon detection probabilities. 

We identify again the parameters of Eq.~(\ref{BWT}) with the following detection probabilities
\begin{eqnarray}\label{set5}
&\hspace{-4mm}p^N(\delta_1,\vartheta_1,\lambda)\!=\!x_1,\;p^N(\delta_2,\vartheta_1,\lambda)\!=\!x_2,\nonumber\\
&\hspace{-6mm}p^N(\delta_3,\vartheta_2,\lambda)\!=\!x_3,\;p^N(\delta_4,\infty,\lambda)\!=\!x_4,
\end{eqnarray}
where $\infty$ indicates once more that the polarization filter is removed for the particular measurement. The constraint $x_4\geq x_3,x_2,x_1$ of the inequality~(\ref{BWT}) is thus guaranteed by the no-enhancement condition~\cite{Clauser:1969:a,Clauser:1974:a,Ou:1988:a}: the detection probability with a polarization filter cannot exceed the measurement without a polarization filter. Following the usual 

\begin{widetext}

assumptions of an LHV theory, we define the joint detection probability exactly as in Eq.~(\ref{pro8}). Using this relation together with~(\ref{set5}), the inequality~(\ref{BWT}), after multiplying by $g(\lambda)$ and integrating over $\lambda$, reads
\begin{eqnarray}\label{BWT2}
T_N&:=&\left[p^N_{12}(\delta_1,\delta_4;\vartheta_1,\infty)-p^N_{12}(\delta_1,\delta_2;\vartheta_1,\vartheta_1)\right.\nonumber\\
&-&\left.p^N_{12}(\delta_1,\delta_3;\vartheta_1,\vartheta_2)+p^N_{12}(\delta_2,\delta_3;\vartheta_1,\vartheta_2)\right]/p^2_{12}(\delta_1,\delta_2;\infty,\infty)\geq0.\hspace{1cm}
\end{eqnarray}
In analogy to the foregoing Section and to provide a better comparability with the results obtained so far we normalized Eq.~(\ref{BWT2}) again by the factor $p^2_{12}(\delta_1,\delta_2;\infty,\infty)$ which is independent of $N$ (c.f.~Eq.~(\ref{pro7c})). In the following, we refer to the inequality~(\ref{BWT2}) as {\em homogeneous Bell-Wigner} (HBW) inequality.

\end{widetext}

\subsection{Violation of the HBW inequality for a system of multiple single-photon emitters\label{42}}

In this subsection, we will test the HBW inequality~(\ref{BWT2}) for $N\geq2$ single-photon emitters by determining the minimum values of $T_N$. For this purpose, employing Eqs.~(\ref{e555}),~(\ref{e556}) and~(\ref{e557}), we make use of the joint detection probabilities Eqs.~(\ref{pro7a}) - (\ref{pro7c}) and search for the minima of $T_N$. However, as the analyses get involved and analytically intricate, we only provide numerical results, which we obtained by scanning through the complete parameter space of $\delta_1,\delta_2,\delta_3,\delta_4$ for each $N$ separately.

\begin{figure}[b!]
\centering
\includegraphics[width=0.45\textwidth, bb=60 440 500 830, clip=true]{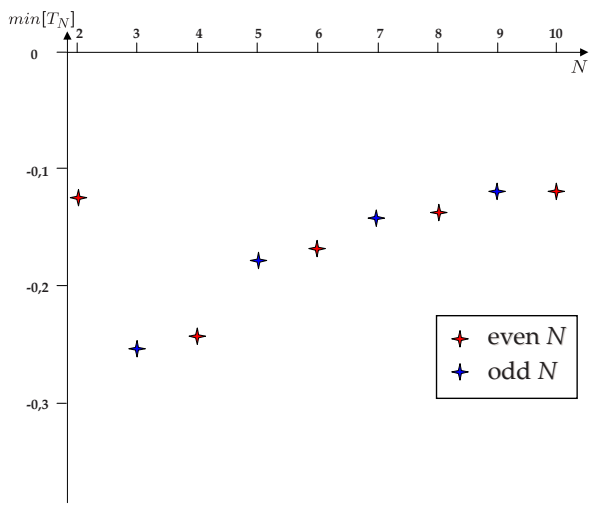}
\caption{\label{f58} Minimal values of $T_N$ (c.f.~Eq.~(\ref{BWT2})) for $N=2,...,10$. The plot illustrates a steady violation of the HBW inequality Eq.~(\ref{BWT2}).}
\end{figure}

The results for the minima of $T_N$ ($min[T_N]$) for $N=2,...,10$ are displayed in Fig.~\ref{f58}. It shows that we have $min[T_N] < 0$ for $N=2,...,10$. For $N=3$ we obtain the lowest value of $min[T_3]\approx-0.254$ and for $N=10$ we have $min[T_{10}] \approx-0.118$. Even though the values of $min[T_N]$ increase monotonously for $N>2$, numerical calculations indicate that they approach zero only for $N\rightarrow\infty$. This suggests that a violation of the HBW inequalities~(\ref{BWT2}) can be obtained for any finite $N$. 

Note that for $N=2$ we obtain $min[T_2]=-0.125$ which sticks out of the overall behavior. We explain this outlier by the fact that $T_2$ depends only on three of the four parameters $\delta_1,\delta_2,\delta_3,\delta_4$ since, due to destructive interference, we have $p^2_{12}(\delta_1,\delta_4;\vartheta_1,\infty)=\frac{{\cal C}_0^2}{4}$, i.e., a constant independent of $\delta_1$ and $\delta_4$. In contrast, for $T_N$ with $N>2$ the term $p^N_{12}(\delta_1,\delta_4;\vartheta_1,\infty)$ is not a constant and thus can be employed to shift $min[T_N]$ towards smaller values. 



\subsection{Interrelationsship between violation of the HBW inequality and  visibility of the two-photon correlation signal\label{43}}

Let us again consider the theoretically attainable visibility ${\cal V}_N$ of the two-photon correlation signal $G^{(2)}_N(\delta_1,\delta_2;\frac{\pi}{4},\frac{\pi}{4})$ (c.f~Fig.~\ref{f57}). From Eq.~(\ref{VN}) we know that it is given by ${\cal V}_N=\frac{N}{3\,N-4}$ which reaches 50\% for $N=4$ and approaches 33\% for $N\rightarrow\infty$. At the same time we can see from Fig.~\ref{f58} that the HBW inequality remains continuously violated when increasing the number of emitters $N$. This shows that the HBW inequality Eq.~(\ref{BWT2}) can be violated by an intensity correlation signal of second order $G^{(2)}_N(\delta_1,\delta_1,\frac{\pi}{4},\frac{\pi}{4})$ having a visibility of below 71\%. In fact, for finite $N$, our results show that a system of $N$ regularly arranged single photon emitters always displays spatial correlations among the scattered photons which violate the criterion of locality even though the visibility of the two-photon correlation signal approaches 33\%. This clearly demonstrates that without indicating N the magnitude of the visibility of the G$^{(2)}$-signal can not be taken as a signature of non-locality.

\section{Conclusion\label{5}}

In conclusion, we investigated the non-local behavior of a system of N $\geq$ 2 particles, i.e., photons emitted by a chain of N independent single photon emitters. Path entanglement among the emitted photons is created  in the process of detection due to the absence of which-way information when registering a photon in the far field of the source. Introducing a new homogenous Bell-Wigner inequality and employing simple photon-photon correlation functions which are experimentally easily implementable in the laboratory we showed that this inequality can be violated for any finite number $N$ even though the visibility of the two-photon signal approaches 33\% in this case. The violation of the homogenous Bell-Wigner inequality unambigiously proves the non-local correlations of the emitted particles. In contrast, using the well-known CH74 inequalities, it turned out that no such violation can be obtained for $N>2$. For this a visibility greater than 71\% is required which cannot be achieved for $N>2$.

\section{Acknowledgements}

C.T.~and J.v.Z.~gratefully acknowledge financial support by the Staedtler foundation. R.W.~thanks the Elite Network of Bavaria for funding.

\appendix
\section{Proof of inequality~(\ref{BWT})\label{app2}}

In this appendix we prove the inequality~(\ref{BWT}) which is an extension of the Bell-Wigner inequality~(\ref{pit}) (see, e.g., \cite{Pitowsky:1989:a}). The Bell-Wigner inequality usually reads
\begin{eqnarray}
0\leq x_1-x_1\,x_2-x_1\,x_3+x_2\,x_3,
\end{eqnarray}
which is valid under the condition $0\leq x_1,x_2,x_3\leq1$.

As explained in Sec.~\ref{4}, it is advantageous to use the inequality~(\ref{BWT}) which reads
\begin{eqnarray}\label{BWTA}
0\leq x_1\,x_4-x_1\,x_2-x_1\,x_3+x_2\,x_3,
\end{eqnarray}
consisting of products of the form $x_i\,x_j$ ($i,j=1,2,3,4$) only. Eq.~(\ref{BWTA}) holds if $1\geq x_4\geq x_1,x_2,x_3\geq0$ is fulfilled.

For the proof we consider two cases:\vspace{2mm}\\
First, we assume $x_2\geq x_1$. In this case we can rewrite the inequality~(\ref{BWTA}) as
\begin{eqnarray}
0\leq x_1\,(x_4-x_2)+x_3\,(x_2-x_1),
\end{eqnarray}
which is valid since both brackets are positive or zero due to the fact that $x_2\geq x_1$ and $x_4\geq x_2$ (note that $x_1,x_2,x_3,x_4\geq0$).

Second, we assume $x_1>x_2$. Here, we make a further case differentiation: let us assume $x_1\geq x_3$ and rewrite the inequality~(\ref{BWTA}) as 
\begin{eqnarray}
0\leq x_1\,(x_4-x_2)-x_3\,(x_1-x_2).
\end{eqnarray}
This inequality is valid since the first bracket is bigger or equals the second due to $x_4\geq x_1$ and since $x_1\geq x_3$. In contrast, if we assume $x_3>x_1$, we can rewrite the inequality~(\ref{BWTA}) as
\begin{eqnarray}
x_1\,(x_4-x_3)-x_2\,(x_1-x_3)\stackrel{x_3>x_1}{>}x_1\,(x_4-x_3)\geq0,
\end{eqnarray}
where the last inequality holds due to $x_4\geq x_3$.

\end{document}